 \documentclass[authoryear,preprint,review,12pt]{elsarticle}

\usepackage{amssymb}

\def\astrobj#1{#1}
\journal{New Astronomy}

\begin{document}

\begin{frontmatter}



\title{New $\beta$ Cephei variable in the Southern open cluster \astrobj{NGC 6200}:  \astrobj{ALS 3728}}


\author[l1]{C. Ulusoy\corref{cr}}
\author[l2]{E. Niemczura}
\author[l3]{B. Ula{\c s}}
\author[l1]{T. G{\"u}lmez}
\address[l1]{Department of Physics, University of Johannesburg, P.O. Box 524, APK Campus, 2006, Johannesburg, South Africa}
\address[l2]{Astronomical Institute of Wroc\l{}aw University, ul. Kopernika 11, 51-622 Wroc\l{}aw, Poland}
\address[l3]{Department of Astronomy and Space Sciences, University of Ege, 35100, Bornova, {\.I}zmir, Turkey}
\cortext[cr]{Corresponding author. Tel.:+90 538 4263084. \\
E-mail address: cerenuastro@gmail.com}

\begin{abstract}
We report the evidence of a new $\beta$ Cephei type variable located in the Southern open cluster \astrobj{NGC 6200}: \astrobj{ALS 3728}. Spectroscopic and photometric observations of NGC\,6200 stars were carried out at the South African Astronomical Observatory (SAAO) in 2011. It was found that \astrobj{ALS 3728} has a frequency of about 4.95 d$^{-1}$ with highest amplitude. The star also shows a remarkable stillstand phase just before its light maxima which is a typical characteristic occurring among these type of stars. Furthermore, a mode identification is applied for the dominant frequency that is calculated from the Fourier analysis.
\end{abstract}


\end{frontmatter}

\section{Introduction}

$\beta$\,Cephei variables are a well known group of Population I early B-type pulsators. Their pulsations were characterized by mono and multiperiodic variations with periods between 2 and 8 hours \citep{sta05}. They are massive (8-18 $M_\odot$) near main sequence stars \citep{sta05} and pulsate both in $p$ (accoustic-pressure) and $g$ (gravity) modes caused by $\kappa$ mechanism through the elements of iron group \citep{cox92,dzi93a,dzi93b}.

Recently, a number of short period variables were reported in the large scale surveys by \cite{poj02,poj03,nar06,pig05,kol04}. In particular, during the third stage of All Sky Automated Survey (ASAS-3), a catalog of over 38000 stars was published by \cite{pig05} and new $\beta$ Cephei stars were defined as members of the Southern young open clusters. According to the survey,  \astrobj{NGC 6200} is one of these clusters where new $\beta$ Cephei stars and canditate ones were discovered. The location of the cluster member $\beta$ Cephei pulsators on the Hertzsprung-Russell Diagram (HRD) can be used to obtain crucial information on their evolutionary stage and provides the determination of stellar mass, reddening and distance modulus by means of isochrone fitting method. The Southern open cluster \astrobj{NGC 6200} (age 8.5 Myr, \citep{khar05}) is a good candidate for such studies. Recently, $\beta$ Cephei type pulsations of ALS 3721, which is first observed by ASAS-3 survey \citep{pig05}, has been confirmed by \citep{ulu13}.

For this study, we have selected \astrobj{ALS 3728} (	GSC 08330-02266=\astrobj{NGC 6200} \#3) which shows a very strong candidacy as a $\beta$ Cephei pulsator (Pigulski, private communication), to probe into oscilliation properties in both photometric and spectroscopic view. The star observed by ASAS-3 survey, however, there is no other publication available in the literature to prove its $\beta$ Cephei type behaviour in detail. \astrobj{ALS 3728} is also one of the members of the Southern open cluster \astrobj{NGC 6200} and it is positioned at the Southern constellation Ara situated between Scorpius and Triangulum Australe. The first photoelectric and spectroscopic observations of the 15 stars in NGC 6200 were reported by  \cite{fit77}. They determined the colour excess and the distance modulus as: \($E(B-V)$=0^{m}.63 \pm 0^{m}.07 \), \(11^{m}.9 \pm 0^{m}.2 \) for the cluster respectively.

\section{Spectroscopy}
The spectroscopy of  \astrobj{ALS 3728} was performed on 18 June 2011 with the Robert Stobbie Spectrograph (RSS) at the 11-m Southern African Large Telescope (SALT), Sutherland, South Africa. The RSS instrument reached $R\sim 3000$ spectral resolution for the spectra of \astrobj{ALS 3728}. During the observing run, six spectra were acquired with exposure times set to 100 s and reached signal-to-noise ratio from 60 to 80, in the spectral range from 3925 to 5990 \AA. Reduction and calibration of the data was made by the standard IRAF procedures including the {\tt ONEDSPEC} and the {\tt TWODSPEC} packages, and {\tt CONTINIUUM} package for normalization.

In order to estimate atmospheric parameters of the star we compared the
low-resolution averaged SALT spectrum with theoretical spectra
calculated by Thierry Morel \citep{mor06}. The synthetic spectra
were computed with a hybrid non-LTE approach. Hydrostatic,
plane-parallel, line-blanketed LTE atmosphere models were calculated
with the ATLAS\,9 code \citep{kur93}, while a constant microturbulent
velocity, $\xi_{\rm mod}$ = 8~kms$^{-1}$ was adopted. The NLTE population
numbers and synthetic spectra were computed using an updated version of
DETAIL \citep{gid81,but84} and SURFACE \citep{but85}
line-formation codes. DETAIL solves the radiative
transfer and statistical equilibrium equations, while SURFACE calculates
the emergent spectrum. The grid of synthetic spectra was calculated for
effective temperatures between 15000 and 30000\,K with a step equal to
1000\,K and surface gravities ranging from 3.0 to 4.5\,dex with a step
equal to 0.1\,dex. The analysis was based on the spectra computed with
solar chemical composition and with microturbulence velocity equal to 8
\,km\,s$^{-1}$.

\begin{figure*}
\begin{center}
\includegraphics[scale=0.7]{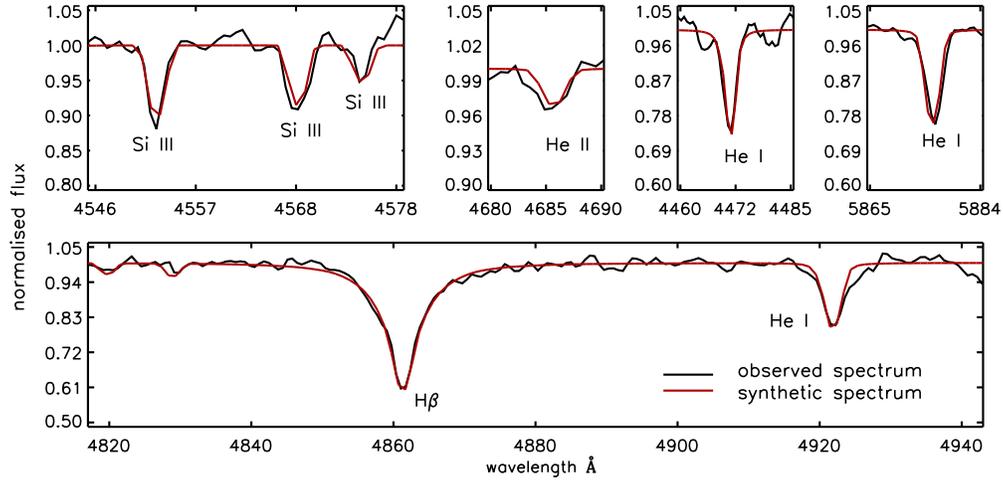}
\caption{Observed (black) and synthetic (red-fitted) spectra of ALS 3728 for selected region. The observed lines fitted to model fluxes with $T_{\rm eff} =25000\pm2000$\,K and $\log g = 3.5\pm0.2$}
\label{figspectrum}
\end{center}
\end{figure*}

The determination of surface gravity $\log g$ and effective temperature
$T_{\rm eff}$ was based on one Balmer line (H$\beta$), He\,I lines
(4471, 4921 and 5876\,\AA), one He\,II line (4686\,\AA) and silicon
lines Si\,III (4552, 4568, 4575\,\AA). Although the quality of the spectrum and signal-to-noise ratio were low the calculations were good enough to estimate the parameters. The best fit of the model fluxes
to the observed lines was obtained as a result of minimizing residual
standard deviation of the fit. The analysis resulted in $T_{\rm eff} =
25000\pm2000$\,K and $\log g = 3.5\pm0.2$. Additionally, the projected
rotational velocity $v \sin i$, was derived to be equal to $100 \pm 20
$\,km\,s$^{-1}$. The results of the spectroscopic analysis are plotted in Fig. \ref{figspectrum}.

\section{Photometry}
\begin{figure}
\includegraphics[scale=0.5]{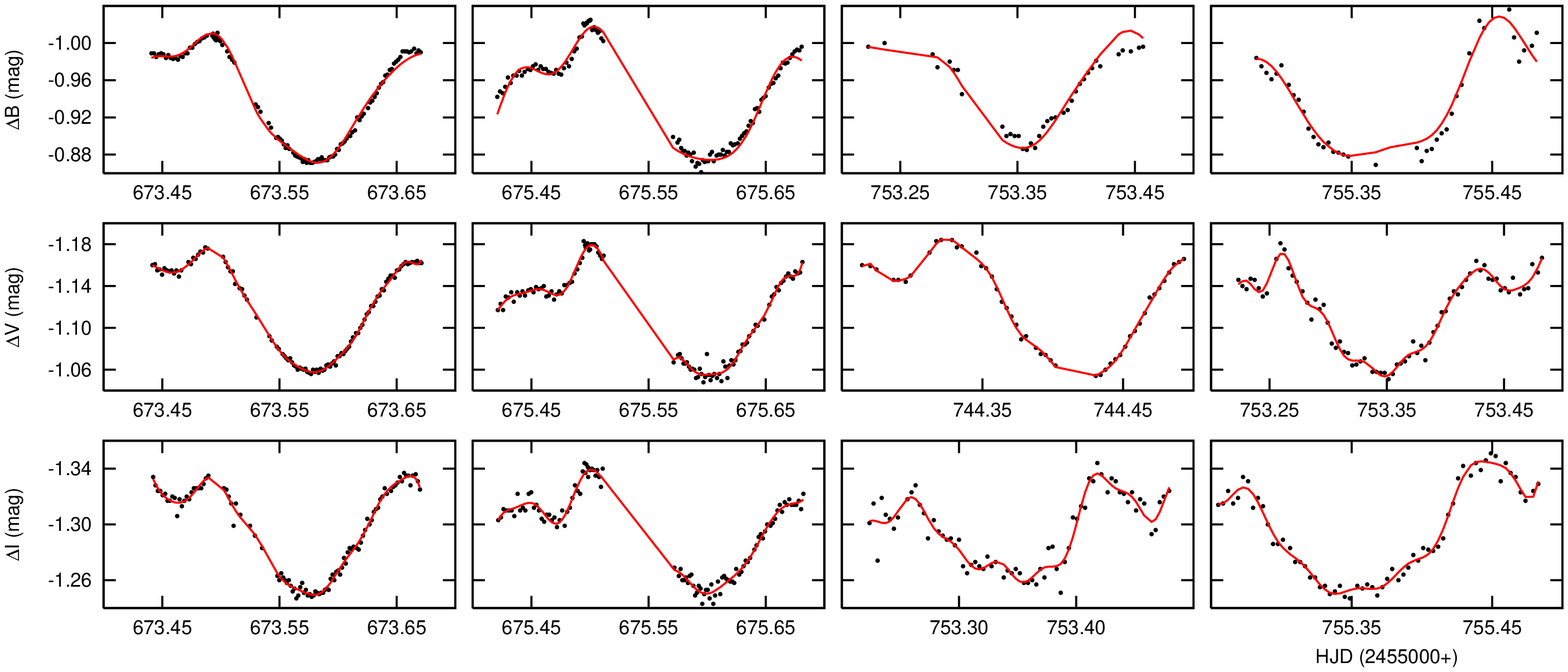}
\caption{The agreement between synthetic and observational light curves of \astrobj{ALS 3728}. Solid lines are determined by frequency analysis.}
\label{figlc}
\end{figure}

CCD photometry of \astrobj{ALS 3728}  was obtained by using the 100-cm Elizabeth Cassegrain telescope at SAAO between the Julian dates 2455673.442-2455755.483. During the observations, a total of 311, 330, and 344 data points in the Johnson {\it{B}}, {\it{V}} and {\it{I}} filters were collected, respectively. CCD data reduction was performed following standard {\tt IRAF} routines including the substraction of  bias and dark frames and flat-field correction for each image. In order to derive instrumental magnitudes of \astrobj{ALS 3728} in the CCD field, aperture photometry technique was used by the {\tt DAOPHOT II} package \citep{ste87}. As a comparison star,GSC 08330-02221  were used. The photometric light curves  in the $BVI$ filters is shown in Fig.~\ref{figlc}. The peak-to-peak amplitude of light variation of the star {bf is equal to} 0$^m$.16 in the B filter.

\begin{figure}
\begin{center}
\includegraphics{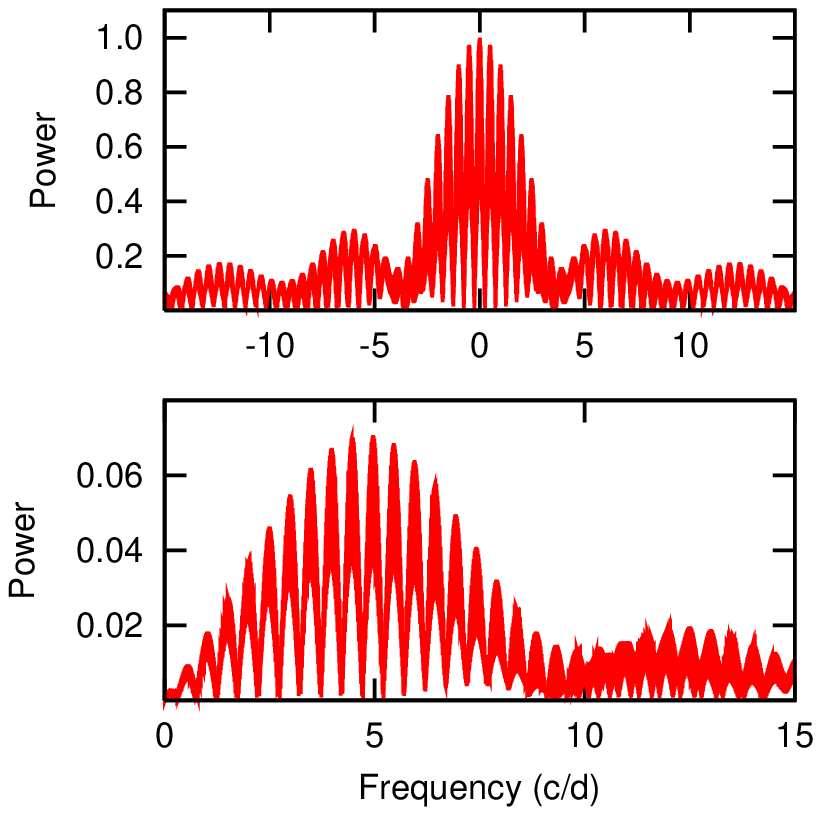}
\caption{The top panel shows the spectral window of the data. Frequency spectra of \astrobj{ALS 3728} is plotted in the bottom panel.}
\label{figspec}
\end{center}
\end{figure}

In order to search for the variability period of the star, we used {\tt PERIOD04} \citep{l05} software package. The package fits the observational data by using the least-square method and is based on classical Fourier analysis. The signal--to--noise ratio threshold was adopted to 4 \citep{bre00}. We proceed the analysis between 0 and 15 c/d limits following \cite{bal11}. The analysis of \astrobj{ALS 3728} allowed us the determination of two dominant pulsation frequencies with strong signals (Table~\ref{tabfr}). Although several other frequencies were found for each filter individually, the large day gap in our observations was attributed as the cause of the presence of those frequencies. An additional frequency $f_{3}$=9.97 d$^{-1}$  can also be considered, however it can only be reached in the $V$ and $I$ filters. As can be seen from the Fig. \ref{figspec}, the highest peaks occurr around 5 d$^{-1}$. Table~\ref{tabfr} lists the result of independent frequencies that were reached in all filters and their amplitudes, phases and signal--to--noise ratios. The agreement of the solution with observational data is shown in Fig.~\ref{figlc}. Fig.~\ref{figspec} illustrates the spectral window and the amplitude spectra of $B$--filter data before prewhitening of any frequencies. We conclude that \astrobj{ALS 3728} shows $\beta$ Cephei type pulsational variability.

\begin{table}
\centering
\caption{Frequencies (d$^{-1}$ ), amplitudes, $A$ (mmag), and phases $\phi$ (radians) determined. Signal-to-noise ratio per frequency is given in the last column. The standard deviations in the last digits are given in parentheses.}
\label{tabfr}
\begin{tabular}{lrrrr}
\hline
\multicolumn{1}{c}{ID} &
\multicolumn{1}{c}{Freq} &
\multicolumn{1}{c}{$A_B$} &
\multicolumn{1}{c}{$\phi_B$} &
\multicolumn{1}{c}{SNR} \\
\hline
$f_1$  & 4.96399(5)      & 0.05947(47) & 0.86020(125)    &139\\
$f_2$  &13.43545(28)      & 0.01139(47) & 0.11137(652)    &29\\
\hline
\multicolumn{1}{c}{ID} &
\multicolumn{1}{c}{Freq} &
\multicolumn{1}{c}{$A_V$} &
\multicolumn{1}{c}{$\phi_V$} &
\multicolumn{1}{c}{SNR} \\
\hline
$f_1$  & 4.92624(4)   &0.05325(38) & 0.67982(114)  &214\\
$f_2$  & 13.43887(24)   &0.01096(38) & 0.72779(563)  &14\\
\hline
\multicolumn{1}{c}{ID} &
\multicolumn{1}{c}{Freq} &
\multicolumn{1}{c}{$A_I$} &
\multicolumn{1}{c}{$\phi_I$} &
\multicolumn{1}{c}{SNR} \\
\hline
$f_1$  & 4.95182(11)   &0.03758(55) & 0.87679(249)  &42\\
$f_2$  & 13.41234(42)   &0.00826(55) & 0.35894(1005)  &6\\

\\
\hline
\end{tabular}
\end{table}

\section{Mode Identification}
\begin{figure*}
\begin{center}
\includegraphics[scale=1.3]{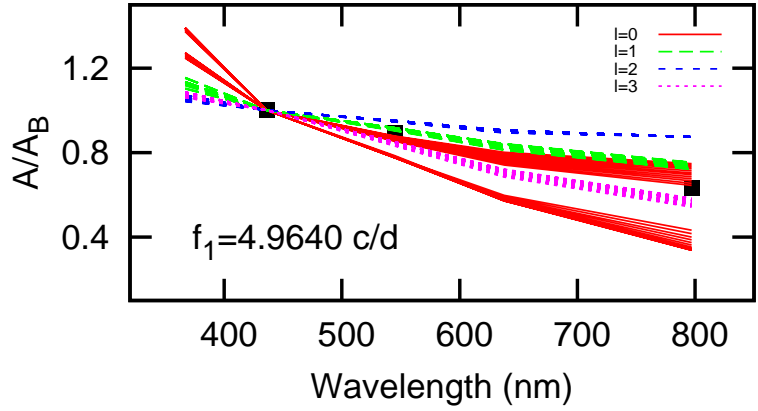}
\caption{Results of the mode identification process. Amplitude ratios are given as a function of the central wavelength normalized to the Johnson B filter for $f_1$. The coloured lines in the panels represent different spherical harmonic degrees corresponding to different stellar models  in the range of  $22500 < T_{\rm eff} < 27500$~K and $3.4 < \log g < 3.8$. The red (solid) lines indicate $l=0$ , the green (long-dashed) $l=1$, the dark blue (dashed) $l=2$ the purple (dotted) $l=3$ . The models computed a mass of 15 M$_\odot$. The black crosses represent the observed values with their 1$\sigma$ standard deviations for $f_1$.}
\label{figmodeid}
\end{center}
\end{figure*}

The multicolour photometry was used to perform mode identification of \astrobj{ALS 3728}. Despite the inedequate number of data points, we attempt to identify the spherical harmonic degree ($l$) for the dominant pulsation frequency of $f_{1}$=4.96 d$^{-1}$. For this purpose, we used the  {\tt FAMIAS} software package \citep{zim08} adopting the $1.6 < M_\odot < 20$, ZAMS to TAMS and A04, {\tt OPAL}\footnote{{\tt OPAL} opacities computed with non-adiabatic Warsaw-New Jersey/Dziembowski code by J. Daszy{\'n}ska-Daszkiewicz and P. Walczak ({\it http://helas.astro.uni.wroc.pl}).}. For the process of mode identification we follow the steps which are explained in detail by \cite{ulu13}. The theoretical amplitude ratios of different degrees $l$ were computed in the range of $22500 K < T_{\rm eff} < 27500 K$ , $3.4<$log g$ < 3.8$, and microturbulance 8 km s$^{-1}$ for the best solution. The results obtained with the stellar mass 15$M_\odot$ which is very close to the value of 15.18$\pm$0.17$M_\odot$ yielded by means of the formulae given by \cite{tor10} which gives mass and radius in terms of $T_{\rm {eff}}$, $\log g$ and $Z$. The amplitude ratios normalized to B filter for the single mode are shown in Fig. \ref{figmodeid}.

\section{Conclusion}

The first evidence of the $\beta$ Cephei type variability of the star \astrobj{ALS 3728} is given in this study. To study seismic behavior of the star located in NGC 6200, we performed photometric and spectroscopic observations. Frequency analysis resulted with the two genuine frequencies, $f_{1}$=4.96 d$^{-1}$  and $f_{2}$=13.43d$^{-1}$. These frequencies refer that the star pulsates in the $\beta$ Cephei type range and are the first determined frequencies in the literature. In the light curves, a stillstand phase occurred just before the maximum brightness (Fig.~\ref{figlc}). This phenomenon is one of the common characteristics observed among some $\beta$ Cephei stars\citep{jer84,ste86,pig02,ulu12}. The effect can be observed mostly during the pulsation period of the star before the maxima and it takes several minutes to half and hour. BW Vul is well known and well studied target among $\beta$ Cephei variables that also shows a remarkable stillstand phase \citep{jer84,ste86}in the light curve. Several possible scenarios were suggested by many authors in the literature to explain this interesting feature. The one accepted for BW Vul is that the bump of the light curve caused by gas dynamics occured between the two strong shock waves \citep{fok04,smi03,mat98}. The origin of first shock is connected to the inner parts of the star where the $\kappa$ mechanism drives by the iron bump elements. It then brakes before reaching the stellar surface by the second shock that is originated from outer layers. In other words,  the stillstand can be explained as the result of complicated atmosphere motions involved by the two shock waves.

Spectroscopic observations and analysis indicate that \astrobj{ALS 3728} has a high rotational velocity ( $v \sin i$=$100 \pm 20$\,km\,s$^{-1}$) with derived effective temperature $T_{\rm eff} = 25000\pm2000$\,K and $\log g = 3.5\pm 0.2$. Using the distance modulus of \astrobj{NGC 6200}, $m-M=13^{m}.36$, given by {\tt WEBDA}\footnote{{\it http://www.univie.ac.at/webda/webda.html}} database the luminosity of the star is found to be log${L \over L_{\odot}}$=3.59. The apparent visual magnitude of the star was taken from {\tt SIMBAD}\footnote{{\it http://simbad.u-strasbg.fr/simbad/}} database and and the solar absolute magnitude was adopted from \cite{cox00}.

Using the stellar parameters in the given range( 15$M_\odot$ $22500 K < T_{\rm eff} < 27500 K$ , logg $3.4<$log g$ < 3.8$) mode identification was carried out to determine the degree $l$  for at least the primary oscillation mode of \astrobj{ALS 3728}.  Our results showed that the amplitude ratios for $f_{1}$=4.96 d$^{-1}$ are closely represented with the theoretical amplitude ratios between the values of $l=0$ and $l=1$. Due to the lack of observational data points, we can not reach a definite conclusion about the identification and further investigation is needed. However, the primary oscillation mode of \astrobj{ALS 3728} might be associated with the fundamental radial mode which could be mostly expected in these type of variables \citep{daz02}.

All in all, $\beta$ Cephei members of open clusters play an important role in order to study their evolutionary state that can be found by using colour-magnitude diagrams. They indeed need necessary attention to understand the nature of their pulsations and possible correlations between pulsational parameters and metal abundance. 

\section*{Acknowledgments}
CU sincerely thanks the South African National Research Foundation (NRF) for the award of innovation post doctoral
fellowship, grant No. 73446. EN acknowledges support from the NCN grant 2011/01/B/ST9/05448. Calculations have been carried out in Wroclaw Centre for Networking and Supercomputing (http://www.wcss.wroc.pl), grant No. 214. This paper uses observations made at the South African Astronomical Observatory (SAAO). Some of observations reported in this paper were obtained with the South African Large Telescope (SALT) carried out with the project ID number 2011-2-RSA\_POL-ULUSOY as CU the P.I. The authors wish to thank SALT astronomer Dr. Petrie Vaisanen and Dr. Alexei Kniazev for helping the spectroscopic observations. The authors also thank Prof. Andrzej Pigulski for sharing information on the target selection. This study was made by using the {\tt WEBDA} database, operated at the Institute for Astronomy of the University of Vienna, the NASA Astrophysics Data System and the SIMBAD database, operated at CDS, Strasbourg, France.

\end{document}